\documentclass[aps,showpacs]{revtex4}
\usepackage{graphicx}
\usepackage{float}
\usepackage[T1]{fontenc}
\usepackage[latin1]{inputenc}
\usepackage{amssymb}
\include{epsf}
\epsfverbosetrue

\newcommand{\xx}{\noindent}

\newcommand{\bea}{\begin{eqnarray}}
\newcommand{\eea}{\end{eqnarray}}

\makeatother
\begin{document}

\title{\Large{Leading Order QED Electrical Conductivity from the 3PI Effective Action}}

\author{M.E. Carrington and E. Kovalchuk}

\email{carrington@brandonu.ca; kavalchuke@brandonu.ca}
 \affiliation{Department of Physics, Brandon University, Brandon, Manitoba, R7A 6A9 Canada\\ and \\
  Winnipeg Institute for Theoretical Physics, Winnipeg, Manitoba }

\begin{abstract}
In this article we study the electrical conductivity in QED using the resummed 3PI effective action. 
We work to 3-loop order in the effective action. We show that the resulting expression for the conductivity is explicitly gauge invariant, and that the integral equations that resum the pinching and colinear contributions are produced naturally by the formalism. All leading order terms are included,  without the need for any kind of power counting arguments.
\end{abstract}

\maketitle
\section{Introduction}

It is well known that selective resummations can play an important role in quantum field theory. A prominant example is the hard thermal loop theory, which includes screening effects that regulate infra-red divergences. The convergence of such perturbation theories can typically be improved using an $n$-particle irreducible ($n$PI) effective action. In addition, $n$PI approximation schemes are of particular interest because they can be used to study far from equilibrium systems. 

The $n$PI effective action is consistent with the global symmetries of the theory, however, the ward identities may not be satisfied during intermediate steps of the  calculation \cite{AS,HZ}. To address this problem we look at the resummed $n$PI effective action, which is defined with respect to the self consistent solutions of the $n$-point functions. This type of strategy was originally proposed by Baym and Kadanoff \cite{baym} and has been discussed in the context of scalar theories in Ref. \cite{vanh}. The renormalization of resummed theories has been discussed in Ref. \cite{reinosa}. The resummed effective action respects all symmetry properties of the theory, and the $n$-point functions which are obtained by functional differentiation satisfy the ward identities. In this paper we study the resummed qed 3PI effective action. We demonstrate how the calculation of transport coefficients is organized in the framework of this formalism.

`Pinch singularities' play an important role in the calculation of transport coefficients. 
The basic idea is that there is an 
infinite number of terms that all contribute at the same order because of the low frequency limit in the kubo formula (Eqn. (\ref{cond})). This limit produces pairs of retarded and advanced propagators which carry the same momenta. When integrating a term of the form $\int dp_0 \;G^{ret}(P)G^{adv}(P)$, the integration contour is `pinched' between poles on each side of the real axis, and the integral contains a divergence called a `pinch singularity.' These divergences are regulated by using resummed propagators which account for the finite width of thermal excitations. This procedure introduces extra factors of the coupling in the denominators which change the power counting. As a consequence, there is an infinite set of graphs which contain products of pinching pairs that all need to be resummed. This set of graphs is resummed by solving an integral equation whose kernel is the square of the matrix elements that correspond to the 2 $\rightarrow$ 2 scattering and production processes. 

For the qed electrical conductivity, it has already been demonstrated that this integral equation can be obtained from the 2PI formalism. In Ref. \cite{gert3} it was shown that the 2-loop contribution to the 2PI effective action produces the square of the $s$-channel, which gives the complete result at the leading-log order of accuracy. In a previous paper \cite{MC-EK}, we have shown that the 3-loop terms in the 2PI effective action contribute the missing $t$- and $u$- channels so that the full matrix element corresponding to all binary scattering and production processes is obtained.

However, the result obtained in Ref. \cite{MC-EK} is not complete at leading order. 
In gauge theories, in addition to pinch singularities, the presence of collinear singularities makes the 1 $\rightarrow$ 2 scatterings as important as the 2 $\rightarrow$ 2 scatterings. These collinear terms must also be resummed by solving an integral equation. 
The complete leading order result is then obtained by solving these two coupled integral equations.
In this paper we derive both of these integral equations directly from the 3-loop resummed 3PI effective action. Thus we show that the full leading order contribution to the qed electrical conductivity can be obtained directly from the 3PI formalism. 

Our result agrees with that obtained previously in \cite{AMY}. This calculation is not obtained directly from quantum field theory but is derived from kinetic theory.  In addition, the conductivity has recently been calculated using a diagramatic method based on power counting arguments in which the ward identity is used explicitly to select contributions that will produce a gauge invariant result \cite{jeon2}. 
Using our method, gauge invarience is automatically satisfied, and all leading order terms appear without the need for any kind of power counting arguments. 
The method we develop in this paper should be generalizable to the calculation of other transport coefficients, like the shear viscosity.

Our calculation provides a field theoretic connection to the kinetic theory results of \cite{AMY}, which is useful in itself. In addition, it seems likely that quantum field theory provides a better framework than kinetic theory for calculations beyond leading order. Our results provide strong support for the use of $n$PI effective theories as a method to study the equilibration of quantum fields. 

This paper is organized as follows. 
In section \ref{Notation} we define some notation. 
In section \ref{3PI} we review the 3PI formalism. 
In section \ref{tildeGamma} we discuss the resummed 3PI formalism. In \ref{EoM} we give the equations of motion, in \ref{bS} we derive the corresponding bethe-salpater equations, and in \ref{Dext} we obtain an expression for the external photon propagator that satisfies the usual ward identity.
In section \ref{Cond} we use these results to obtain an expression for the conductivity that is complete to leading order.  
In section \ref{Conclusions} we present our conclusions and discuss future directions. 

\section{Notation}
\label{Notation}

The electrical conductivity can be obtained from the kubo formula:
\bea
\label{cond}
&&\sigma=\frac{1}{6}\left(\frac{\partial}{\partial q_0}2\, {\rm Im}\,\rho^{ii}(q_0,0)\right)\Big|_{q_0\rightarrow 0}\\[2mm]
&&\rho^{ii}(x,y)=\langle j^i(x)j^i(y)\rangle\,;~~j^i(x)=\bar\psi(x)\gamma^i\psi(x)\nonumber
\eea
We can write the conductivity in terms of the polarization tensor using:
\bea
\label{cond2}
\rho^{ii}(q_0,0)=-\frac{1}{e^2}\Pi^{ii}_{ret}(q_0,0)
\eea
In the rest of this paper, the coupling constant $e$ will be absorbed into the definition of the vertex functions, and not written explicitly.

\begin{figure}[H]
\begin{center}
\includegraphics[width=5cm]{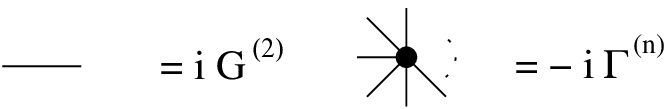}
\end{center}
\caption{Definitions of notation for propagators and vertices.}
 \label{idefn2}
\end{figure}

In order to simplify the notation we will use a single numerical subscript to represent all continuous and discrete indices. For example: a photon field is written $A_1:=A^a_{\mu}(x)$; 
the fermion propagator is written $S_{12}:=S_{\alpha\beta}^{ab}(x_1,x_2)$, etc.
We also use an einstein convention in which a repeated index implies a sum over discrete variables and an integration over space-time variables. We define $2$-point green functions and $n$-point vertex functions as shown in Fig. \ref{idefn2}.
\par

\section{The 3PI formalism}
\label{3PI}

The 3PI effective action to 3-loop order can be written \cite{berges}:
\begin{eqnarray}
\label{Gamma3PI}
&&\Gamma[\psi, \bar\psi, A, S, D,V,U]\\
&&~~~~=S_{cl}[\psi, \bar\psi, A]+
    \frac{i}{2} {\rm Tr} \,{\rm Ln}D^{-1}_{12}+
\frac{i}{2} {\rm Tr}\left[(D^0_{12})^{-1}\left(D_{21}-D^0_{21}\right)\right]-
  i {\rm Tr} \,{\rm Ln} S^{-1}_{12} - 
i{\rm Tr} \left[(S^0_{12})^{-1}(S_{21}-S^0_{21})\right]\nonumber\\
&&~~~~+\Gamma_2^0[S,D,V,U]++\Gamma_2^{\rm int}[S,D,V,U] \nonumber
\eea
where $S$ is the fermion propagator, $D$ is the photon propagator, $V$ is the fermion-photon vertex, $U$ is the 3-photon vertex (which is zero at the tree level), $S_{cl}[\psi, \bar\psi, A]$ is the classical action, and $S_0$ and $D_0$ are the free propagators given by:
\bea
\label{S0D0}
(S^0_{12})^{-1}=\frac{\delta^2S_{cl}}{\delta\psi_2\delta\bar\psi_1}\,;~~(D^0_{12})^{-1}=\frac{\delta^2 S_{cl}}{\delta A_2\delta A_1}\,.
\eea
The functions $\Gamma_2^0[S,D,V,U]$ and $\Gamma_2^{\rm int}[S,D,V,U]$ are shown graphically in Fig. \ref{3piGamma}.
\par\begin{figure}[H]
\begin{center}
\includegraphics[width=12cm]{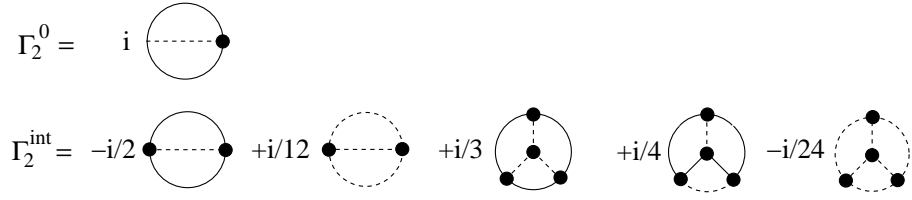}
\end{center}
\caption{3PI effective action to 3-loop order.}
 \label{3piGamma}
\end{figure}
\noindent The first diagram in Fig. \ref{3piGamma} contains a bare fermion-photon vertex which is defined as
\bea
V^0_{132}=i\frac{\delta^3 S_{cl}}{\delta\psi_2\delta A_3 \delta\bar\psi_1}
\eea

The equations of motion of the mean fields, propagators and vertices are obtained from the stationarity of the action:
\bea
\label{eom1}
&&\frac{\delta \Gamma[\psi, \bar{\psi},A,S,D,V,U]}{\delta A}=0\,;~~\frac{\delta \Gamma[\psi, \bar{\psi},A,S,D,V,U]}{\delta \psi}=0\,;~~\frac{\delta \Gamma[\psi, \bar{\psi},A,S,D,V,U]}{\delta \bar\psi}=0\\
&&\frac{\delta \Gamma[\psi, \bar{\psi},A,S,D,V,U]}{\delta S}=0\,;~~\frac{\delta \Gamma[\psi, \bar{\psi},A,S,D,V,U]}{\delta D}=0\nonumber\\
&&\frac{\delta \Gamma[\psi, \bar{\psi},A,S,D,V,U]}{\delta V}=0\,;~~\frac{\delta \Gamma[\psi, \bar{\psi},A,S,D,V,U]}{\delta U}=0\nonumber
\eea

\section{The 3PI Resummed Effective Action}
\label{tildeGamma}

The last four equations in (\ref{eom1}) can be solved simultaneously for the self-consistent solutions: $\tilde{S}[\psi,\bar{\psi},A]$, $\tilde{D}[\psi,\bar{\psi},A]$, $\tilde{V}[\psi,\bar{\psi},A]$ and $\tilde{U}[\psi,\bar{\psi},A]$ which satisfy:
\bea
\label{eom2}
&&~~~ \frac{\delta \Gamma[\psi, \bar{\psi},A,S,D,V,U]}{\delta S}
  \bigg|_{\tilde{S},~\tilde{D},~\tilde{V},~\tilde{U}} = \frac{\delta \Gamma[\psi, \bar{\psi},A,S,D,V,U]}{\delta D}
  \bigg|_{\tilde{S},~\tilde{D},~\tilde{V},~\tilde{U}}\\
  &&= \frac{\delta \Gamma[\psi, \bar{\psi},A,S,D,V,U]}{\delta V} \bigg|_{\tilde{S},~\tilde{D},~\tilde{V},~\tilde{U}} = \frac{\delta \Gamma[\psi, \bar{\psi},A,S,D,V,U]}{\delta U}
  \bigg|_{\tilde{S},~\tilde{D},~\tilde{V},~\tilde{U}} = 0\nonumber
\eea
Substituting the self consistent solutions we 
obtain the resummed action:
\begin{eqnarray}
\label{Gamma3PI-rs}
&& \tilde{\Gamma}[\psi,\bar{\psi},A]=
  \Gamma[\psi,\bar{\psi},A,\tilde{S}[\psi,\bar{\psi},A],\tilde{D}[\psi,\bar{\psi},A],\tilde{V}[\psi,\bar{\psi},A],\tilde{U}[\psi,\bar{\psi},A]]
\end{eqnarray}
The equivalence of (\ref{Gamma3PI}) and (\ref{Gamma3PI-rs}) at the exact level was shown in \cite{CJT}.

\subsection{Constraint Equations}
\label{EoM}

Now we look at the equations of motion (\ref{eom2}). 
The first two equations have the form of dyson equations:
\bea
\label{dyson}
&& \tilde S_{12}^{-1} = (S^0_{12})^{-1} - \Sigma_{12}\,;~~~~\Sigma_{12} = -i \frac{\delta \Gamma_2[S, D,V,U]}{\delta S_{21}}\Big|_{\tilde S,~\tilde D,~\tilde V,~\tilde U}\\
&& \tilde D_{12}^{-1} = (D^0_{12})^{-1} - \Pi_{12}\,;~~~~\Pi_{12} = 2 i \frac{\delta \Gamma_2[S,D,V,U]}{\delta D_{21}}\Big|_{\tilde S,~\tilde D,~\tilde V,~\tilde U}\nonumber
\eea
where we have defined $\Gamma_2=\Gamma_2^0+\Gamma_2^{\rm int}$. We substitute (\ref{Gamma3PI}) into (\ref{dyson}) and take the derivatives. The results are given in Eqn. (9).  For the fermion-photon vertex the order of indices is always (outgoing fermion, photon, incoming fermion). In each term, the factors that carry fermion indices are written in square brackets,  and the order of the indices reflects the flow of fermion number. The order of the indices in the photon propagator and the 3-photon vertex does not matter, because of the symmetry of these $n$-point functions. The results are:
\bea
\label{dyson2}
&&\tilde S_{12}^{-1} = (S^0_{12})^{-1}-2i [\tilde V_{13'4'}\tilde S_{4'4}V^0_{432}]\cdot \tilde D_{33'}+i [\tilde V_{13'4'}\tilde S_{4'4}\tilde V_{432}]\cdot \tilde D_{33'} \\[4mm]
&&~~~~ +[\tilde V_{15'6'}\tilde S_{6'6}\tilde V_{63'7'}\tilde S_{7'7}\tilde V_{754'}\tilde S_{4'4}\tilde V_{432}]\cdot \tilde D_{55'}\tilde D_{33'}
+[\tilde V_{14'7'}\tilde S_{7'7}\tilde V_{75'6'}\tilde S_{6'6}\tilde V_{632}]\cdot \tilde U_{3'45}\tilde D_{33'}\tilde D_{44'}\tilde D_{55'}
\nonumber\\[4mm]
&&\tilde D_{12}^{-1} = (D^0_{12})^{-1}
+2i [\tilde V_{413'}\tilde S_{3'3}V^0_{324'}\tilde S_{4'4}] - i [\tilde V_{413'}\tilde S_{3'3}\tilde V_{324'}\tilde S_{4'4}]
+\frac{i}{2}\tilde U_{13'4}\tilde U_{234'}\tilde D_{3'3}\tilde D_{4'4} \nonumber\\[4mm]
&&~~~~ - [\tilde V_{514'}\tilde S_{4'4}\tilde V_{47'3'}\tilde S_{3'3}\tilde V_{326'}\tilde S_{6'6}\tilde V_{675'}\tilde S_{5'5}]\cdot \tilde D_{77'}
-2[\tilde V_{514'}\tilde S_{4'4}\tilde V_{43'7'}\tilde S_{7'7}\tilde V_{765'}\tilde S_{5'5}]\cdot \tilde U_{236'}\tilde D_{33'}\tilde D_{66'}\nonumber\\[4mm]
&&~~~~+\frac{1}{2}\tilde U_{14'5}\tilde U_{3'47'}\tilde U_{5'67}\tilde U_{236'}\tilde D_{3'3}\tilde D_{4'4}\tilde D_{5'5} \tilde D_{6'6}\tilde D_{7'7}\nonumber
\eea
These equations are shown in Fig. \ref{dysonFig}. 
\par\begin{figure}[H]
\begin{center}
\includegraphics[width=12cm]{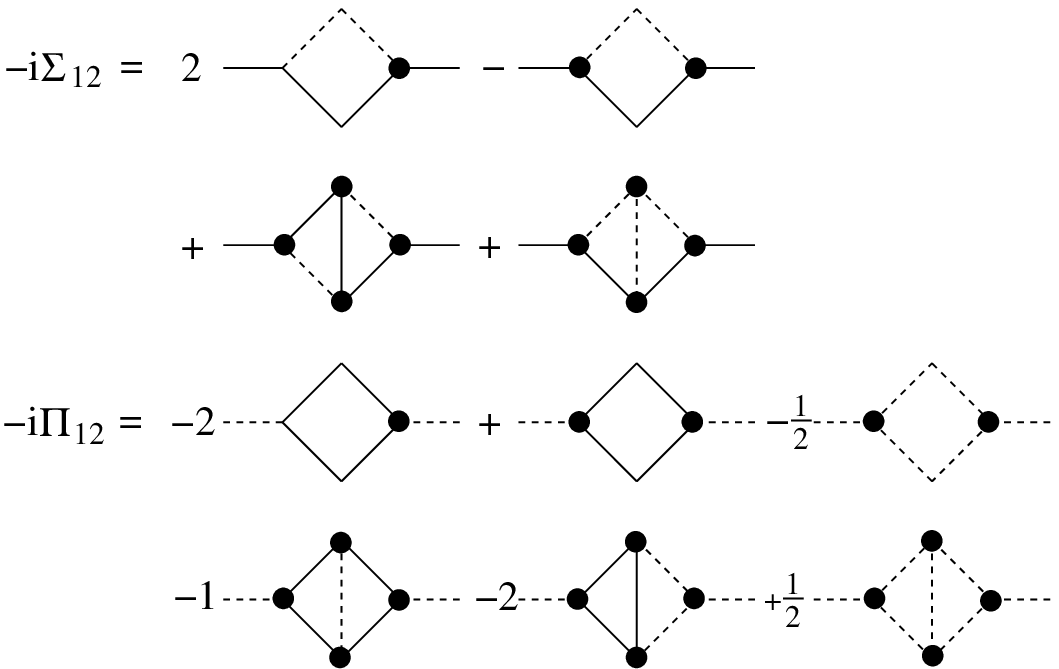}
\end{center}
\caption{Self energy and polarization tensor.}
 \label{dysonFig}
\end{figure}

Using (\ref{Gamma3PI}) the second two equations in (\ref{eom2}) can be written:
\bea
\label{UVeqn}
&& \tilde V_{132}=V^0_{132}+i \tilde V_{175'}\tilde S_{5'5}\tilde V_{536}\tilde S_{66'}\tilde V_{6'7'2}\tilde D_{7'7}+i \tilde V_{75'1}\tilde D_{5'5}\tilde U_{536}\tilde D_{66'}\tilde V_{26'7'}\tilde S_{7'7}\\[2mm]
&&\tilde U_{132}=-2i \tilde V_{715'}\tilde S_{5'5}\tilde V_{536}\tilde S_{66'}\tilde V_{6'27'}\tilde S_{7'7}+i \tilde U_{715'}\tilde D_{5'5}\tilde U_{536}\tilde D_{66'}\tilde U_{6'27'}\tilde D_{7'7}\nonumber
\eea
These equations can be represented graphically as shown in Fig. \ref{UVfig}.
\par\begin{figure}[H]
\begin{center}
\includegraphics[width=8cm]{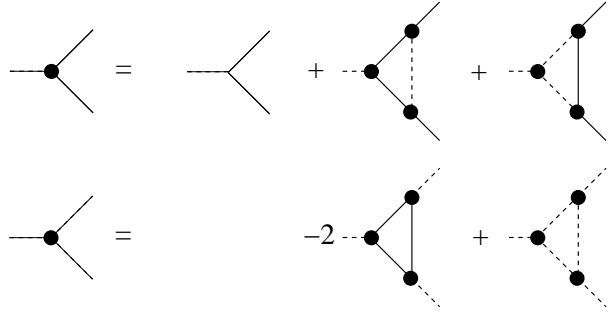}
\end{center}
\caption{Self-consistent equations for the vertices.}
 \label{UVfig}
\end{figure}

\subsection{Bethe-Salpeter Equations}
\label{bS}

Taking derivatives of the dyson equations we obtain a set of bethe-salpeter type integral equations. In order to write these equations in a simple way, we define the following vertex functions:
\begin{eqnarray}
\label{vert-defns}
&& \Lambda^0_{132} = -\frac{\delta (S^{0}_{12})^{-1}}{\delta A_3}\,;~~\Lambda_{132} = -\frac{\delta \tilde{S}^{-1}_{12}}{\delta A_{3}}\,;~~\Omega_{132} = -\frac{1}{2}\frac{\delta \tilde{D}^{-1}_{12}}{\delta A_{3}}
\end{eqnarray}
These functions are always written with the external photon on the middle leg, and this external photon is distinguished by an arrow in figures. To clarify the notation, all of the types of vertices that we use in this paper are shown in Fig. \ref{vertFig}.
\par\begin{figure}[H]
\begin{center}
\includegraphics[width=12cm]{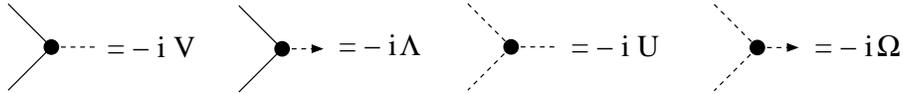}
\end{center}
\caption{The vertices.}
 \label{vertFig}
\end{figure}
Some additional useful relations can be obtained from the identities:
\begin{eqnarray}
\label{in0}
\tilde{S}^{-1}_{13}\tilde{S}_{32}=\delta_{12}\,;~~\tilde{D}^{-1}_{13}\tilde{D}_{32}=\delta_{12}
\end{eqnarray}
Differentiating (\ref{in0}) with respect to $A$ and using (\ref{vert-defns}) gives:
\bea
\label{invert}
&& \frac{\delta\tilde{S}_{12}}{\delta A_{3}}= \tilde S_{11'}\Lambda_{1'3 2'}\tilde{S}_{2'2}\,;~~
\frac{\delta\tilde{D}_{12}}{\delta A_{3}} = 2
\tilde{D}_{11'}  \Omega_{1'32'}\tilde{D}_{2'2}
\eea

The bethe-salpeter equations are obtained by differentiating (\ref{dyson}) and using the definitions (\ref{vert-defns}) to obtain:
\bea
\label{bS1}
&& \Lambda_{132}=\Lambda^0_{132}+\frac{\delta\Sigma_{12}}{\delta A_3}\;;~~~~ \Omega_{132}=\frac{1}{2}\;\frac{\delta\Pi_{12}}{\delta A_3}
\eea
where $\Sigma_{12}$ and $\Pi_{12}$ are obtained from Eqns (\ref{dyson}) and (\ref{dyson2}) (and shown in Fig. \ref{dysonFig}). 
When taking derivatives, numerous cancellations occur between contributions from different pieces of the self energy. To show how these cancellations occur, we distinguish two different kinds of propagators: `rung propagators' do not connect, at either end, to a vertex that attaches to an external leg, and non-rung propagators do connect to a vertex that attaches to an external leg. For example, the third diagram in Fig. \ref{dysonFig} gives a contribution to $\Sigma$ that has one fermion rung propagator, two fermion non-rung propagators and two photon non-rung propagators. After differentiating and collecting terms, the result is that 
all 2-loop terms containing derivatives of non-rung propagators cancel. 
We illustrate this point by looking at contributions to $\delta\Sigma_{12}/\delta A_3$ containing derivatives of non-rung photon propagators. We give the results in terms of diagrams only. Differentiating the first two contributions to $\Sigma$ in Fig. \ref{dysonFig} and using (\ref{invert}) gives the terms shown in Fig. \ref{sigma1}.
\par\begin{figure}[H]
\begin{center}
\includegraphics[width=7cm]{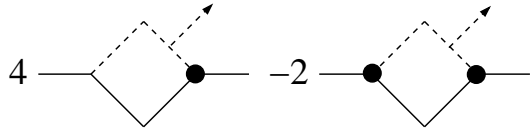}
\end{center}
\caption{Result from differentiating non-rung photon propagators in the first two contributions to $-i\Sigma$ in Fig. \ref{dysonFig}.}
 \label{sigma1}
\end{figure}
\noindent We rewrite this result by substituting in for $V_0$ using Eqn. (\ref{UVeqn}) (shown in Fig. \ref{UVfig}) and obtain the terms shown in Fig. \ref{sigma11}.
\par\begin{figure}[H]
\begin{center}
\includegraphics[width=12cm]{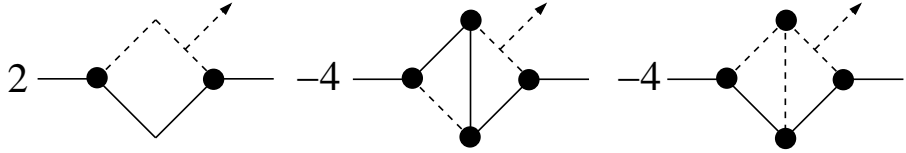}
\end{center}
\caption{Result after substituting in for $V_0$.}
 \label{sigma11}
\end{figure}
\noindent Differentiating the third and fourth contributions to $-i\Sigma$ in Fig. \ref{dysonFig} we obtain the terms shown in Fig. \ref{sigma2}.
\par\begin{figure}[H]
\begin{center}
\includegraphics[width=7cm]{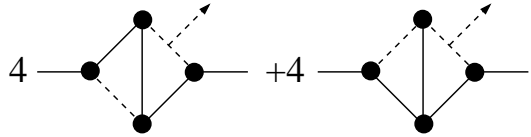}
\end{center}
\caption{Result from differentiating non-rung photon propagators in the second two contributions to $-i\Sigma$ in Fig. \ref{dysonFig}.}
 \label{sigma2}
\end{figure}
\noindent Combining the results shown in Figs. \ref{sigma11} and \ref{sigma2}  we see that all 2-loop terms containing derivatives of non-rung photon propagators cancel. The same effect occurs for derivatives of non-rung fermion propagators. In Fig. \ref{lambdaEqn} we give the result for the first equation in (\ref{bS1}) to 2-loop order and in Fig. \ref{omegaEqn} we give the result for the second equation in (\ref{bS1}) to 1-loop order. In section \ref{Cond} we will show that higher loop terms do not contribute to the conductivity at leading order.
\par\begin{figure}[H]
\begin{center}
\includegraphics[width=15cm]{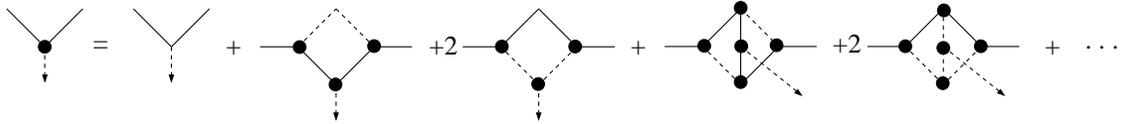}
\end{center}
\caption{Bethe-salpeter equantion for the $\Lambda$ vertex to 2-loop order.}
 \label{lambdaEqn}
\end{figure}
\par\begin{figure}[H]
\begin{center}
\includegraphics[width=10cm]{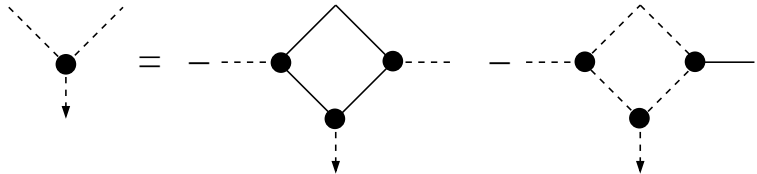}
\end{center}
\caption{Bethe-salpeter equantion for the $\Omega$ vertex to 1-loop order.}
 \label{omegaEqn}
\end{figure}

\subsection{External Propagator}
\label{Dext}

The external photon propagator is defined as 
\bea
\label{ext-prop}
(D^{{\rm ext}}_{12})^{-1}=   
     \frac{\delta^2}{\delta A_2 \delta A_1}
     \tilde{\Gamma}[\psi,\bar{\psi},A]
\eea
In this section we show that the propagator defined in (\ref{ext-prop}) satisfies the usual ward identities. The basic mechanism is simple:
the dyson equations (\ref{dyson}) contain $s$-channel ladder resummations and the bethe-salpeter equations (\ref{bS1}) introduce $t$- and $u$-channels, and thus restore the crossing symmetry.

We start by taking the derivative of the modified effective action using the chain rule. In the rest of this section we suppress the arguments and write $\Gamma[\psi,\bar\psi,A, S, D,V,U]$ as $\Gamma$. We use the notation $X_i$ to indicate one of the set $X:=\{S,D,V,U\}$ and $\tilde X_i$ to indicate one of the set $\tilde X:=\{\tilde S,\tilde D,\tilde V,\tilde U\}$. The sum $\frac{1}{2}\sum_{i\ne j}$ gives the 6 terms in the set of pairs $X_i\ne X_j$ where order is not respected. We write the result:
\begin{eqnarray}
\label{Dext-long}
(D^{{\rm ext}}_{12})^{-1} 
&&= \frac{\delta^2\Gamma}{\delta A_{2} \delta A_{1}}\Big|_{\tilde X}
+\sum_{i}\frac{\delta^2\Gamma}{\delta X_i^2}\Big|_{\tilde X}\frac{\delta \tilde X_i}{\delta A_2}\frac{\delta \tilde X_i}{\delta A_1}\\
&&+\Big[\sum_{i}\frac{\delta^2\Gamma}{\delta X_i\delta A_1}\Big|_{\tilde X}\frac{\delta \tilde X_i}{\delta A_2}+\frac{1}{2}\sum_{i\ne j}\frac{\delta^2\Gamma}{\delta X_i\delta X_j}\Big|_{\tilde X}\frac{\delta \tilde X_i}{\delta A_1}\frac{\delta \tilde X_j}{\delta A_2}~~+~~\{1\leftrightarrow 2\}\Big]\nonumber
\end{eqnarray}

This expression can be further simplified by using the set of equations obtained by differentiating the equations of motion:
\bea
\frac{\delta}{\delta A}\;\Big[\frac{\delta \Gamma}{\delta X_i}\Big|_{\tilde X}\Big]=0~~\Rightarrow~~
\frac{\delta^2 \Gamma}{\delta X_i\delta A}\Big|_{\tilde X}+\sum_{j}\frac{\delta^2\Gamma}{\delta X_j\delta X_i}\Big|_{\tilde X}\frac{\delta \tilde X_j}{\delta A} = 0
\eea
We solve these four equations for the four derivatives:
$\delta^2\Gamma/\delta X_i^2|_{\tilde X}$
and substitute into (\ref{Dext-long}) to obtain:
\bea
(D^{{\rm ext}}_{12})^{-1} =\frac{\delta^2\Gamma}{\delta A_{2} \delta A_{1}}\Big|_{\tilde X}+\frac{1}{2}\Big[\sum_{i}\frac{\delta^2\Gamma}{\delta X_i\delta A_1}\Big|_{\tilde X}\frac{\delta \tilde X_i}{\delta A_2}~~+~~\{1\leftrightarrow 2\}\Big]
\eea
Expanding the sum and using (\ref{invert}) we have:
\bea
(D^{{\rm ext}}_{12})^{-1} &&=\frac{\delta^2\Gamma}{\delta A_{2} \delta A_{1}}+\frac{1}{2}\Big[
\frac{\delta^2\Gamma}{\delta S_{34}\delta A_1}\Big|_{\tilde X}\cdot(\tilde S_{33'}\Lambda_{3'24'}\tilde S_{4'4})
+2\frac{\delta^2\Gamma}{\delta D_{34}\delta A_1}\Big|_{\tilde X}\cdot(\tilde D_{33'}\Omega_{3'24'}\tilde D_{4'4}) \\[2mm]
&&+\frac{\delta^2\Gamma}{\delta U_{345}\delta A_1}\Big|_{\tilde X}\frac{\delta \tilde U_{345}}{\delta A_2}
+\frac{\delta^2\Gamma}{\delta V_{345}\delta A_1}\Big|_{\tilde X}\frac{\delta \tilde V_{345}}{\delta A_2}
~~+~~\{1\leftrightarrow 2\}\Big]\nonumber
\eea
Using (\ref{Gamma3PI}) we obtain:
\bea
\frac{\delta^2\Gamma}{\delta A_{2} \delta A_{1}} &=& (D_{12}^0)^{-1}\\
\frac{\delta^2\Gamma}{\delta S_{34}\delta A_1}&=&-i\frac{\delta (S^0_{34})^{-1}}{\delta A_1} = i \Lambda^0_{314} \nonumber\\
\frac{\delta^2\Gamma}{\delta D_{34}\delta A_1}&=&\frac{\delta^2\Gamma}{\delta V_{345}\delta A_1}=\frac{\delta^2\Gamma}{\delta U_{345}\delta A_1}=0 \nonumber
\eea
which gives the final result:
\bea
\label{dext}
(D^{{\rm ext}}_{12})^{-1}=(D^{0}_{12})^{-1}+i \Lambda^0_{314}\cdot[\tilde S_{44'}\Lambda_{4'23'}\tilde S_{3'3}]
\eea
This expression has the form of the usual schwinger-dyson equation. 
Following the method of \cite{MC-EK} it is straightforward to show that the external photon propagator in (\ref{dext}) satisfies the ward identity. 

From Eqns. (\ref{dyson}) and (\ref{dext}) we extract the vertex part of the 2-point function:
\bea
\label{piFirst}
\Pi^{{\rm ext}}_{12}=-i(\Lambda^{0}_{314}\tilde{S}_{44'}\Lambda_{4'23'}\tilde{S}_{3'3})
\eea
This result is illustrated in Fig. \ref{piF}. 
\par\begin{figure}[H]
\begin{center}
\includegraphics[width=5cm]{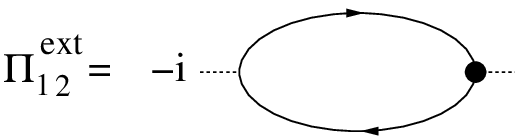}
\end{center}
\caption{Graphical representation of Eqn (\ref{piFirst}).}
 \label{piF}
\end{figure}

\section{Leading Order Conductivity}
\label{Cond}

The conductivity is obtained from the polarization tensor through Eqns. (\ref{cond}) and (\ref{cond2}).
The polarization tensor depends on the propagator $\tilde S$ and the vertex $\Lambda$ through Eqn. (\ref{piFirst}).
The two equations in (\ref{dyson2}), the two equations in (\ref{UVeqn}), and the equations represented in Figs. \ref{lambdaEqn} and \ref{omegaEqn}, form a set of six coupled integral equations for the six quantities $\{\tilde S,~\tilde D,~\tilde V,~\tilde U,~\Lambda,~\Omega\}$. In this section we show that these equations contain all of the leading order contributions to the conductivity. 

The pinch singularities are resummed by an integral equation of the form:
\bea
\label{bs8}
\Lambda_{132} &&=\Lambda^{0}_{132}~~+ \sum_{j\in \{a,b,c,d,e,f,g\}}i\,M^{(j)}_{12;45}\cdot[\tilde S_{55'}  \,\Lambda_{5'34'}\,\tilde S_{4'4}]
\eea
This equation is shown schematically in Fig. \ref{bSy}. 
\par\begin{figure}[H]
\begin{center}
\includegraphics[width=8cm]{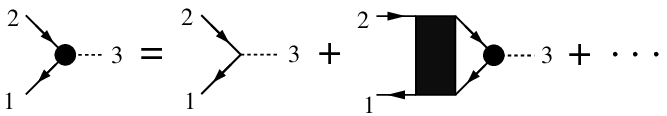}
\end{center}
\caption{Graphical representation of Eqn (\ref{bs8}).}
\label{bSy}
\end{figure}
\noindent The functions $M^{(j)}_{12;45}$ represent 4-point functions where the variables on each side of the semicolon indicate legs that join pinching pairs of propagators. We show below that the seven terms included in the sum over $j\in \{a,b,c,d,e,f,g\}$ correspond to the seven diagrams in Fig. \ref{bSfinal}. 
We start from the equation for $\Lambda$ shown in Fig. \ref{lambdaEqn} and substitute iteratively the equations for $\tilde D$, $\tilde S$, $\tilde U$, $\tilde V$ and $\Omega$, collecting all 2-loop contributions. We describe this process step by step. 

Substituting the equation for $\Omega$ (Fig. \ref{omegaEqn}) into the equation for $\Lambda$ (Fig. \ref{lambdaEqn}) and keeping terms up to 2-loop order, we obtain the result shown in Fig. \ref{lambdaEqn2}. In Fig \ref{lambdaEqn3} we redraw the figure in a more conventional form.
\par\begin{figure}[H]
\begin{center}
\includegraphics[width=11cm]{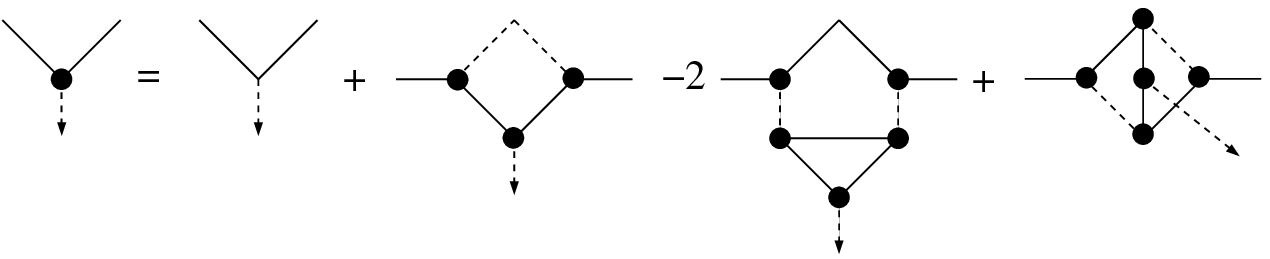}
\end{center}
\caption{Equation for $\Omega$ substituted into the equation for $\Lambda$.}
 \label{lambdaEqn2}
\end{figure}
\par\begin{figure}[H]
\begin{center}
\includegraphics[width=14cm]{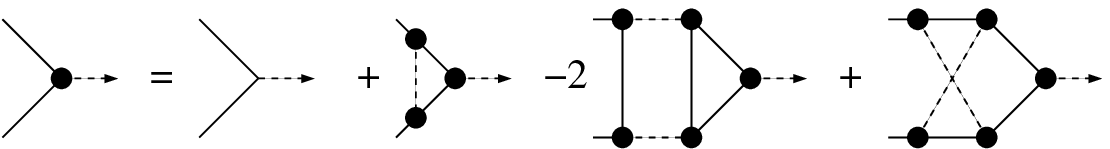}
\end{center}
\caption{Fig. \ref{lambdaEqn2} redrawn.}
 \label{lambdaEqn3}
\end{figure}

The third and fourth diagrams on the right hand side of (\ref{lambdaEqn3}) are already 2-loop order. The fourth diagram is shown in part (c) of Fig. \ref{bSfinal}. We rewrite the third diagram as the sum of the two terms shown in parts (f) and (g) of Fig. \ref{bSfinal}. Note that in Fig. \ref{bSfinal} we include some arrows to indicate the direction of the flow of fermion number, in addition to the arrow on the external photon line associated with the $\Lambda$ vertex. 
It is important to consider terms (f) and (g) separately, in order to show that the correct matrix element is produced by the pinching part of the kernel of the integral equation. The procedure is described in detail in Ref. \cite{MC-EK}.  

The second diagram on the right hand side of Eqn. (\ref{lambdaEqn3}) is of 1-loop order. 
Substituting the equation for $\tilde D$ produces the graph in part (a) of Fig. \ref{bSfinal}. Substituting the equation for $\tilde V$ produces the graphs in parts (d) and (e) of Fig. \ref{bSfinal}.
Iterating the 1-loop term with itself produces the 2-loop diagram shown in part (b) of Fig. \ref{bSfinal}.

\par\begin{figure}[H]
\begin{center}
\includegraphics[width=12cm]{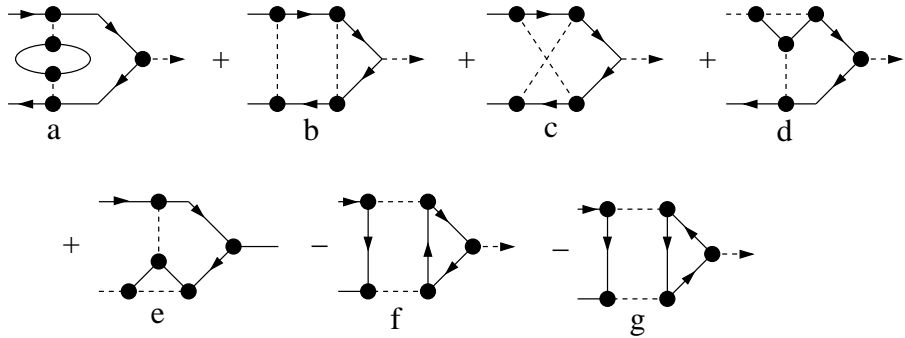}
\end{center}
\caption{2-loop contributions to $\Lambda$.}
 \label{bSfinal}
\end{figure}
Equation (\ref{bs8}) (as shown in Fig. \ref{bSfinal}) is the same as the equation found in Ref. \cite{MC-EK} using the 2PI formalism (see Fig. 8 of Ref. \cite{MC-EK}), except that the vertices are now determined self-consistently through the integral equation (\ref{UVeqn}). The pair of equations (\ref{bs8}) (Fig. \ref{bSfinal}) and (\ref{UVeqn}) (Fig. \ref{UVfig}) are precisely the two equations shown in Fig. 11 of Ref. \cite{jeon2}, and obtained previously, using kinetic theory, in Ref. \cite{AMY}.

\section{Conclusions}
\label{Conclusions}

We have shown that the 3-loop resummed 3PI effective action contains all of the leading order contributions to the conductivity. The two integral equations that resum pinch and colinear terms are produced naturally by the formalism, independently of any power counting analysis. The result is explicitly gauge invariant. The method we develop in this paper should be generalizable to the calculation of other transport coefficients, like the shear viscosity.

Our calculation provides a connection between $n$PI effective theories and kinetic theories, and supports the use of $n$PI effective theories as a method to study the equilibration of quantum fields. \\

\xx This work was supported by the Natural Sciences and Engineering Research Council of Canada.

\newpage

\end{document}